\pdfoutput=1
\documentclass[english]{new_tlp}
\usepackage[latin1]{inputenc}
\usepackage[T1]{fontenc}
\usepackage{babel}
\usepackage{graphicx} 
\usepackage{mathptmx}
\usepackage{amsmath}
\usepackage{amsfonts}
\usepackage{amssymb}
\usepackage{multirow}
\usepackage{stmaryrd}
\usepackage{lmodern}  
\usepackage{xspace}
\usepackage{enumitem}
\usepackage{hyperref}
\usepackage{color}
\usepackage{soul}
\usepackage{breakcites}
\usepackage{tikz}

\definecolor{gcommentcolor}{rgb}{0, 0.6, 0}

\newcommand{\If}{\leftarrow}
\newcommand{\Imp}{\rightarrow}

\newcommand{\eq}{\!=\!}
\newcommand{\diff}{\!\neq\!}
\newcommand{\gr}{\!>\!}
\newcommand{\greq}{\!\geq\!}
\newcommand{\plus}{\!+\!}
\newcommand{\less}{\!<\!}
\newcommand{\lesseq}{\!\leq\!}
\newcommand{\minus}{\!-\!}

\newcommand{\vars}{\mathit{vars}}

\newcommand{\false}{\mathit{false}}

\newcommand{\boldbar}{|\hspace{-.9mm}|\hspace{-.9mm}|\hspace{-.9mm}|}

\newcommand{\su}{\hspace{-.2mm}\raisebox{-2pt}{-}\hspace{-.2mm}} 
\newcommand{\difflist}{\mathit{diff}\!\su\hspace{-.2mm}\raisebox{-2pt}{-}\hspace{-.2mm}\mathit{list}} 
 
\newcommand{\minleaf}{{\textit{min}}\hspace{.3mm}\su {\textit{leaf}}}
\newcommand{\leftdrop}{{\textit{left}}\hspace{.3mm}\su {\textit{drop}}}

\newcommand{\verimap}{{\small\textsc{VeriMAP}}\xspace}
\newcommand{\zthree}{\textsc{Z3}\xspace}
\newcommand{\rcaml}{{\small\textsc{RCaml}}\xspace}
\newcommand{\spacer}{\textsc{Spacer}\xspace}

\newcommand{\EC}{{\mathcal {E\hspace*{-.5pt}C}}}
\newcommand{\secs}{{\hspace*{.3mm}s}}

\newcommand{\TreeProcessing}{{\it Tree\su Processing\/}}

\newcommand{\AlgE}{${\mathcal E}$}

\newtheorem{theorem}{Theorem}         
\newtheorem{example}{Example}         
\newtheorem{definition}{Definition}   

\title{Solving Horn Clauses on Inductive Data Types Without Induction}

\author[E. De~Angelis, F. Fioravanti, A. Pettorossi, M. Proietti]
	{EMANUELE DE ANGELIS,~~ FABIO FIORAVANTI\\
	DEC, `G. d'Annunzio' University of Chieti\/-Pescara, Pescara, Italy\\
	\email{\{emanuele.deangelis, fabio.fioravanti\}@unich.it}
	\and
	ALBERTO PETTOROSSI\\
	DICII, University of Rome Tor Vergata, Rome, Italy\\
	\email{pettorossi@info.uniroma2.it}
	\and
	MAURIZIO PROIETTI\\
	CNR-IASI, Rome, Italy\\
	\email{maurizio.proietti@iasi.cnr.it} 
}

\begin{document}
	
	\maketitle

	\begin{abstract}
	We address the problem of verifying the satisfiability
of Constrained Horn Clauses (CHCs) based on theories
of inductively defined data structures, such as lists and trees.
We propose a transformation technique whose objective is
the removal of these data structures from CHCs, hence
reducing their satisfiability to a satisfiability problem
for CHCs on integers and booleans.
We propose a transformation algorithm and identify a class of clauses where 
it always succeeds. We also consider an extension of that
algorithm, which combines clause transformation with reasoning on
integer constraints.
Via an experimental evaluation 
we show that our technique greatly improves the effectiveness of applying the
Z3 solver to CHCs. We also show that our verification
technique based on CHC transformation followed by CHC solving,
is competitive with respect to CHC solvers extended with induction.

\noindent
\textit{This paper is under consideration for acceptance in TPLP.}	
	\end{abstract}

    \begin{keywords}
	Program verification, 
	constrained Horn clauses, 
	constraint logic programming, 
	inductively defined data types, 
	program transformation.
	\end{keywords}
	
	\section{Introduction}	
	\label{sec:Intro}
	Constraint logic programs have become a well-established formalism for solving
program verification problems~\cite{Al&07,Bj&15,De&14b,Ja&11b,Me&07,Pe&98}. 
In the verification field, constraint logic programs are often called
{constrained Horn clauses} (CHCs), and here we will adopt this terminology.
The verification method based on CHCs consists in reducing a program verification problem
to the satisfiability of a set of CHCs.
Since CHC satisfiability, also called CHC {\em solving}, is in general an undecidable problem, 
some heuristics have been proposed in the literature, such as 
{\em Counterexample Guided Abstraction Refinement} (CEGAR)~\cite{Cl&00},
 {\em Craig interpolation}~\cite{McM03}, and  {\em Property Directed Reachability} 
(PDR)~\cite{Bra11,HoB12}.
Some tools, called CHC solvers,
for verifying the satisfiability of CHCs are available. We recall
Eldarica~\cite{Ho&12}, HSF~\cite{Gr&12}, RAHFT~\cite{Ka&16}, 
VeriMAP~\cite{De&14b}, and Z3~\cite{DeB08}.
CHC solvers make use of a combination of the above heuristics
and have been shown to be very effective for CHCs with
several underlying constraint theories, such as the theory of
Linear Integer Arithmetic ($\mathit{LIA}$) and the theory of Boolean 
constraints ({\it Bool}\/).

Solving techniques have also been developed for CHCs
manipulating inductively defined data structures such as 
lists and trees (see, for instance, {\small \url{https://rise4fun.com/Z3/tutorial/guide}} for the Z3
solver). 
However, CHC solvers acting on those data structures are usually less effective 
than CHC solvers for clauses with constraints in $\mathit{LIA}$ or $\mathit{Bool}$.
This is mainly due to the fact that 
methods for satisfiability
used by CHC solvers are based on variants of resolution, augmented with ad hoc
algorithms for the underlying constraint theory, and
no  induction principles for the data structures are used.

To overcome this difficulty, recent work has proposed the extension of CHC
solving by adding a principle of induction on predicate derivations~\cite{Un&17}.
This work is on the same line of other proposals which extend techniques for
{Satisfiability Modulo Theory} (SMT) with 
inductive reasoning~\cite{ReK15,Su&11} by incorporating methods derived from the field of
automated theorem proving.

In this paper we propose an alternative method to solve CHCs on inductively defined
data structures. It is based on the application of suitable transformations
of CHCs that have the objective of removing those data structures 
while preserving satisfiability of clauses.
Our transformation method makes use of the fold/unfold transformation rules~\cite{EtG96,TaS84},
and it is based on some techniques proposed in the past for improving the 
efficiency of functional and logic programs, such as {\em deforestation}~\cite{Wad90},
{\em unnecessary variable elimination}~\cite{PrP95a}, and {\em conjunctive partial deduction}~\cite{De&99}.
However, the focus of the method presented in this paper is on the improvement of effectiveness of
CHC solvers, rather than the improvement of efficiency of programs.

In our method we separate the concern of reasoning on
inductively defined data structures from the concern of proving clause satisfiability. 
This separation of concerns eases the task of the CHC solvers
in many cases.
For instance, when the constraints are on trees of integers,
our transformation method, if successful, allows us to derive an 
equisatisfiable set of clauses with constraints on integers only.

The main contributions of our work are the following.
(1)~We propose a transformation algorithm, called ${\mathcal E}$, that removes inductively 
defined data structures from
	CHCs and derives new, equisatisfiable sets of clauses whose constraints are
	from the $\mathit{LIA}$ theory and the $\mathit{Bool}$ theory (see Section~\ref{sec:Transform}).	
(2)~We identify a class of CHCs where Algorithm~${\mathcal E}$ 
    is guaranteed to terminate, and
    all inductively defined data structures can be removed (see Section~\ref{sec:Correctness}).
(3)~{By using reasoning techniques on $\mathit{LIA}$ constraints, 
    we derive an extension of our transformation 
    algorithm~${\mathcal E}$, called Algorithm~$\EC$ (see Section~\ref{sec:Generalization}).}
(4)~We report on an experimental evaluation of 
    a tool that implements Algorithm~$\EC$. 
    We consider CHCs that encode {properties} of OCaml
	functional programs,
	and we show that our tool is competitive with the \rcaml 
	tool that extends the Z3 solver by adding induction rules for reasoning on inductive data structures~\cite{Un&17}
	(see Section~\ref{sec:Experiments}).


	\section{A Tree Processing Example}
	\label{sec:Intro-Ex}
	As an introductory example to illustrate our technique for proving program properties   
without using induction on data structures, let us consider 
the following 
{\it Tree\su Processing\/} program, 
which we write according to the OCaml syntax~\cite{Le&17}. 

\smallskip
\noindent
{{\textsf{type} \textit{tree} = \textsf{Leaf} \,\boldbar\, \textsf{Node of} \textit{int} $\ast$ \textit{tree} $\ast$ \textit{tree}\,;;\nopagebreak

\noindent
\textsf{let} \textit{min} $x$ $y$ = $\mathsf{if}$ $x\!<\!y$ $\mathsf{then}$ $x$ $\mathsf{else}$ $y$\,;;

\noindent
\textsf{let} \textsf{rec} {\minleaf} $t$ = $\mathsf{match}~t~\mathsf{with}$

\hspace{20mm}
$\boldbar~\textsf{Leaf}~\textsf{-->}~0$ 

\hspace{20mm}
$\boldbar~ \textsf{Node\/}(x,l,r)~\textsf{-->}~1 + \textit{min}~({\minleaf}\ l)~({\minleaf}\ r)$\,;;

\noindent
\textsf{let} \textsf{rec} {\leftdrop} $n$ $t$ = $\mathsf{match}~t~\mathsf{with}$

\hspace{20mm}
$\boldbar~\textsf{Leaf}~\textsf{-->}~\textsf{Leaf}$

\hspace{20mm}
$\boldbar~ \textsf{Node\/}(x,l,r)~\textsf{-->}~~\mathsf{if}$ $n\!<=\!0$ $\mathsf{then}$ $\textsf{Node\/}(x,l,r)$ 
$\mathsf{else}$ ${\leftdrop}~(n\!-\!1)~l$\,;;

}} 

\smallskip
\noindent
Let us also consider the following non-trivial property {\it Prop} to be verified for the {\TreeProcessing} program: 

\smallskip

$\forall n\ \forall t.\  n\!\geq\! 0\  \Rightarrow\ \big(((\minleaf\ (\leftdrop\ n\ t)) + n )  
\geq (\minleaf\ t)\big)$.

\smallskip
\noindent
Now we translate the {\TreeProcessing} program and the property~{\it Prop}
into a set of CHCs that are satisfiable
iff {\it Prop} holds for \mbox{\TreeProcessing}. We get the following set of clauses
(clauses 1--7 for the program and clause~8 for the property): 

\smallskip

\noindent
\makebox[4mm][l]{1.}$\mathit{min}(X,Y,Z) \If X\less Y,\ Z\eq X$

\noindent
\makebox[4mm][l]{2.}$\mathit{min}(X,Y,Z)\If X\greq Y,\ Z\eq  Y$ 

\noindent
\makebox[4mm][l]{3.}$\mathit{\minleaf}\,(\mathit{leaf},M)\If M\eq 0$

\noindent
\makebox[4mm][l]{4.}$\mathit{\minleaf}\,(\mathit{node}(X,L,R),M)\If M\eq M3\plus 1,\ \mathit{\minleaf}\,(L,M1),\ \mathit{\minleaf}\,(R,M2),\ $ 
      
\hspace*{30mm}$\mathit{min}(M1,M2,M3)$
       
\noindent
\makebox[4mm][l]{5.}$\mathit{\leftdrop}(N,\mathit{leaf},\mathit{leaf}) \If$

\noindent
\makebox[4mm][l]{6.}$\mathit{\leftdrop}(N,\mathit{node}(X,L,R),\mathit{node}(X,L,R))\If N\lesseq 0$

\noindent
\makebox[4mm][l]{7.}$\mathit{\leftdrop}(N,\mathit{node}(X,L,R),T)\If N\greq1,\ N1\eq N\minus 1,\ 
\mathit{\leftdrop}(N1,L,T)$

\noindent
\makebox[4mm][l]{8.}$\false\If N\greq 0,\ M\plus N \less K,\ \mathit{\leftdrop}(N,T,U),\  \mathit{\minleaf}\,(U,M),\ \mathit{\minleaf}\,(T,K)$


\smallskip

\noindent

{
In clause~8  the head 
is \textit{false} because the relation~`$<$' in the constraint $M\plus N \less K$ 
occurring in the body is the negation 
of the relation `$\geq$' occurring in the property~{\it Prop}.
}

The Z3 solver is {\it not} able to 
prove the satisfiability of clauses~1--8 
(which amounts to show that $\false$ is not derivable),
because of the presence of variables of type {\it tree}. One can extend the capabilities of Z3 by adding 
induction rules on trees as done in recent 
work~\cite{ReK15,Su&11,Un&17}. Instead, as we propose in this paper,
 we can derive by transformation, starting from clauses~1--8,
a new, equisatisfiable set of clauses without variables of type {\it tree} and whose constraints 
are in \textit{LIA} only, and then we can use CHC solvers, such as Z3, to show
the satisfiability of this new set of clauses.
Starting from clauses~1--8,  our transformation algorithm, whose details are given in
Section~\ref{sec:Transform}, 
produces the following clauses, where $\mathit{min}$ is defined by clauses~1 and~2:


\smallskip
	
{
{\noindent
\makebox[6mm][l]{~9.}$\mathit{new}1(N,M,K)\If M\eq 0,\ K\eq 0$

\noindent
\makebox[6mm][l]{10.}$\mathit{new}1(N\!,M\!,K)\If N\lesseq 0,\ M\eq M3 \plus 1,\ K\eq M,\ \mathit{new}2(M1),\ \mathit{new}2(M2),\ $ 

\hspace*{30mm}$\mathit{min}(M1,M2,M3)$

\noindent
\makebox[6mm][l]{11.}$\mathit{new}1(N,\!M,\!K)\If N\greq 1, N1\eq N\minus 1, K\eq K3 \plus 1,\ \mathit{new}1(N1,\!M,\!K1),\ \mathit{new}2(K2), $ 

\hspace*{30mm}$\mathit{min}(K1,\!K2,\!K3)$
      
\noindent
\makebox[6mm][l]{12.}$\mathit{new}2(M)\If M\eq 0$

\noindent
\makebox[6mm][l]{13.}$\mathit{new}2(M)\If M\eq M3\plus 1,\ \mathit{new}2(M1),\ \mathit{new}2(M2),\ \mathit{min}(M1,M2,M3)$

\vspace{-.7mm} 
\noindent
\makebox[6mm][l]{14.}$\false\If N\greq 0,\ M\plus N \less K,\ \mathit{new}1(N,M,K)$
}}

\smallskip

\noindent
Now, the Z3 solver for CHCs with \textit{LIA} constraints
is able to prove the satisfiability of clauses~1,\,2,\,9--14
without using any induction on trees.
{The details of the transformation from clauses~1--8 to clauses~1,\,2,\,9--14 will be presented 
in Section~\ref{sec:Transform}.}

	\section{Preliminaries}
	\label{sec:Prelim}
	


Let us consider a typed, first order functional language.
We assume that the {\textit{basic types}} of the language are \textit{int\/}, 
for integers, 
and \textit{bool\/}, for booleans. 
We also have \textit{non-basic types}, which are introduced by 
(possibly recursive) type  
definitions such as the one for trees of integers considered in Section~\ref{sec:Intro-Ex}.
The type system of our language can be formally defined as follows. 
(One may consider richer type systems including, for instance,
parameterized types, but this system is sufficient for presenting 
our verification technique.)

\smallskip

\noindent
\hspace{5mm}\makebox[9mm][l]{\textit{Types}}\ $\ni$
\makebox[2mm][l]{$\tau$}~::= ~\textit{int\/} ~|~ \textit{bool\/}  ~|~ $\mathit{ident}$ ~|~ $\tau_{1}\ast\ldots \ast \tau_{k}$ 

\noindent
\hspace{5mm}\makebox[27mm][l]{\textit{Type Definitions\,{\rm :}}} $\mathit{ident} = $
$c_{1}~\textsf{of}~\tau_{1} ~\boldbar~ \ldots ~\boldbar~ c_{n}~\textsf{of}~\tau_{n}$


\smallskip

\noindent
where: (i)~\textit{ident} is a type identifier, 
(ii)~the operator~`$\ast$' builds $k$-tuple types, 
(iii)~$c_{1}$,~$\dots$,~$c_{n}$ are distinct constructors with arity, 
and (iv)~in any expression `$c_{i}~\textsf{of}~\tau_{i}$', if 
$c_{i}$ has arity~\mbox{$k\!>\! 0$,} then $\tau_{i}$ is a $k$-tuple type,
and if~$c_{i}$ has arity 0, then 
`$\textsf{of}~\tau_{i}$' is absent.

Every function 
has a {\em functional type} of the form $\tau_{1}\rightarrow \tau_{2}$, for
some types~$\tau_{1}$ and~$\tau_{2}$, 
modulo the usual isomorphism between
$\tau \rightarrow (\tau'  \rightarrow \tau'')$ and $(\tau \ast \tau')  \rightarrow \tau''$.

In the following definitions: 
(i)~$n$ is an integer, (ii)~$b$ is {\textit{true}} or  {\textit{false}}, (iii)~$x$~is a typed variable, 
(iv)~$c$~is a constructor of arity~$k\ (\geq 0)$, and
(v)~$f$~is a function of arity~$k\ (\geq 0)$ or a primitive operator on integers or booleans
such as: $+$, $\times$, $=$, $\neq$, $\leq$, $\neg$, $\wedge$, and $\Rightarrow$.

\smallskip
\noindent
\makebox[18mm][r]{~\textit{Values}\ $\ni$}
\makebox[2mm][l]{$v$}~::= ~$n$ ~|~ $b$ ~|~ $c\ v_{1}\ldots v_{k}$


\noindent
\makebox[18mm][r]{~\textit{Terms}\  $\ni$}
\makebox[2mm][l]{$t$}~::=  ~$n$ ~|~ $b$ ~|~ $x$ ~|~ $c\ t_{1}\ldots t_{k}$ ~|~ $f\ t_{1}\ldots t_{k}$ ~|~ 
$\mathsf{if}$ $t_{0}$ $\mathsf{then}$ $t_{1}$ $\mathsf{else}$ $t_{2}$ 

\makebox[20mm][l]{}~|~ $\mathsf{let}$ $x=t_{0}$ $\mathsf{in}$ $t_{1}$ ~|~
   $\mathsf{match}~x~\mathsf{with}~\boldbar~\mathit{p}_{1}~\textsf{-->}~t_{1}~\boldbar~\ldots~\boldbar~ 
   \mathit{p}_{n}~\textsf{-->}~t_{n} $ \nopagebreak
   

\noindent
\makebox[21mm][r]{\textit{~Patterns}\ $\ni$}
\makebox[2mm][l]{$p$}~::= ~$c\,(x_{1},\ldots,x_{k})$


\smallskip
\noindent
We assume that all values, terms, and patterns are well-typed. In every \textsf{let-in} term  $x$ is a 
new variable
not occurring in $t_{0}$ and occurring in $t_{1}$.
In every \textsf{match-with} term the patterns $\mathit{p}_{1},\ldots, \mathit{p}_{k}$ 
are pairwise disjoint and exhaustive.  
\noindent 
A \mbox{{\it user-defined function}}~$f$ is defined by:
$\mathsf{let}~\mathsf{rec}~f\ x_{1} \ldots  x_{k} = t$,
\noindent
where the free variables of $t$ are among $\{x_{1},\ldots, x_{k}\}$.

A \textit{program} is a (possibly empty) set of type definitions together with
a set of \mbox{user-defined} functions.
Given a program~$P$ and a term~$t$ without free variables and whose functions are defined in~$P$, 
the value~$v$ of~$t$ using~$P$ is computed according to the 
call-by-value semantics. In this case we write $t \rightarrow_{P} v$.
We say that a program~$P$ {\textit{terminates}} if, for every term $t$ 
without free variables and whose functions are defined in $P$, there exists a value~$v$ such that 
$t \rightarrow_{P} v$.
Given a boolean term $q$ whose free variables are in $\{x_{1},\ldots,x_{n}\}$,
a {\textit{property}} is a universally quantified formula of the form $\forall x_{1},\ldots,x_{n}.\,q$.
Given a program~$P$, we say that a property $\forall x_{1},\ldots,x_{n}.\,q$, 
whose functions are defined in~$P$, holds for~$P$ iff for all values $v_{1},\ldots,v_{n}$, we have that 
$q\, [v_{1},\ldots,v_{n}~/~x_{1},\ldots,x_{n}] \rightarrow_{P} {\mathit{true}}$, where the
square bracket notation is for substitution.

Let us consider a typed first order logic~\cite{End72} which includes: 
(i)~the theory {\textit{LIA}} of the linear integer arithmetic constraints, 
and (ii)~the theory {\textit{Bool}} of boolean constraints.
A {\em constraint} in $\textit{LIA}\cup\textit{Bool\/}$ is any formula in
$\textit{LIA}\cup\textit{Bool\/}$.

An {\it atom} is a formula of the form $q(t_{1},\ldots,t_{m})$,
where~$q$ is a typed predicate symbol not used in
$\textit{LIA}\cup\textit{Bool\/}$, and 
$t_{1},\ldots,t_{m}$ are typed terms made out of variables and constructors.
A~{\it constrained Horn clause}  (or simply, a {\it clause}, or a CHC) is 
an implication of the form  
$A\leftarrow c, B$ (comma denotes conjunction), 
where the conclusion (or {\it head\/}) $A$ is either an atom or \textit{false}, 
the premise (or {\it body\/}) is the conjunction of
a constraint  $c$, and a (possibly empty) conjunction~$B$ of atoms. 
A clause whose head is an atom is called a {\it definite clause},
and a clause whose head is {\it false} is called a {\it goal\/}.
A {\em constrained fact} is a clause of the form $A\leftarrow c$. We will write the 
constrained fact
$A\leftarrow \textit{true}$ also as $A\leftarrow$.
We assume that all variables in a clause are universally quantified in front.
Given a term $t$, by ${\it vars}(t)$ we denote the set of variables occurring in $t$.
{Similarly for the set of 
variables occurring in atoms, or clauses, or sets of atoms, or
 sets of clauses.}
A set $S$ of CHCs is said to be {\em satisfiable} if $S\cup \textit{LIA}\cup \textit{Bool}$
has a model, or equivalently, $S\cup \textit{LIA} \cup \textit{Bool} ~\not\models \textit{false}$.

	
We define a translation \textit{Tr} from any set $P$ of function definitions 
into a set of CHCs by: 
(i)~introducing for each function $f$ of arity $k$, a new predicate, say $\mathit{p}$,
of arity $k\plus 1$, and then
(ii)~providing the clauses for predicate $p$
so that $f(x_1,\ldots,x_k)\!=\!y$ iff $\mathit{p}(x_1,\ldots,x_k,y)$ holds. 
This translation has been applied in Section~\ref{sec:Intro-Ex} 
for generating {\mbox{clauses~1--7}} starting 
from the \textit{Tree\su Processing\/} program.
Primitive operations, such as $+$ and $-$,
are translated in terms of constraints expressed in \textit{LIA}.
Similarly, a property of the form: $\forall x.\,\mathit{r}(x)\Rightarrow \mathit{s}(x)$ is translated 
into a goal (or a set of goals, in general) as indicated in Section~\ref{sec:Intro-Ex} (see goal~8).
 If we denote by ${\mathit{Tr}}(P)$ and 
$\mathit{Tr}(\mathit{Prop})$ the set of CHCs generated from a given
program~$P$ and a given property $\mathit{Prop}$, respectively, we
have the following theorem, where satisfiability is defined within the typed logic 
we consider.

\begin{theorem}Given a program~$P$ that terminates, a property~$\mathit{Prop}$  holds  for $P$ iff 
the set $\mathit{Tr}(P) \cup \mathit{Tr}(\mathit{Prop})$ of clauses is satisfiable.
\end{theorem}


%

	\section{Eliminating Inductively Defined Data Structures}
	\label{sec:Transform}


Now we present an algorithm, called~Algorithm~\AlgE, for 
eliminating from CHCs the predicate arguments that have a non-basic type, such as lists or trees. 
If Algorithm~$\mathcal E$ terminates, then it transforms a set 
of clauses into an equisatisfiable set, 
where the arguments of all predicates have basic types.
The algorithm is based on the fold/unfold strategy for eliminating unnecessary variables
from logic programs~\cite{PrP95a}.

Given two terms $t_{1}$ and $t_{2}$, by $t_{1}\!\preceq\!t_{2}$ and $t_{1}\!\prec\!t_{2}$ we denote the {\em subterm} and 
{\em strict subterm} relation, respectively. 
Given an atom $A$, by ${\it nbargs}(A)$ we denote the set of arguments of non-basic type of~$A$.
By $\mathit{nbvars}(A)$ we denote the set of variables of \mbox{non-basic} type 
occurring in $A$. 
The next definition extends the $\preceq$ and $\prec$ relations to atoms.


\begin{definition}[Atom Comparison]
Given two atoms $A_1$ and $A_2$,  $A_1 \! \preceq\! A_2$ $($or  $A_1 \! \prec\! A_2$$)$ holds if there exist 
$t_1\!\in\! {\it nbargs}(A_1)$ and $t_2\!\in\! {\it nbargs}(A_2)$ such that:
(i)~$\vars(t_1)\cap\vars(t_2)\diff \emptyset$, and (ii)~$t_1\! \preceq\! t_2$ $($or $t_1\! \prec\! t_2$$)$. 
Given a set~$S$ of atoms, an atom $M\!\in\! S$ is {\em strictly maximal} in~$S$
if there exists no atom $A \!\in\! S$ such that $M \! \prec\! A$ 
and there exists an atom $A'\!\in\! S$ such that $A' \! \prec\! M$. 
\end{definition}

{For example, $q([Y|Ys],[~])$ is strictly maximal in the set $\{p([X],\mathit{Ys}),\ q([Y|\mathit{Ys}],[~])\}$.
	Indeed, $\mathit{Ys}\!\prec\! [Y|\mathit{Ys}]$, and hence $p([X],\mathit{Ys})\!\prec\! q([Y|\mathit{Ys}],[~])$, while
	$q([Y|\mathit{Ys}],[~]) \!\not\prec\! p([X],\mathit{Ys})$ (note that $[~]\!\prec\! [X]$, but 
	$\vars([~])\!\cap\! \vars([X])\eq \emptyset$).
	The notion of a strictly maximal atom will be used in the {\it Unfold} procedure of Algorithm~$\mathcal E$
	to guide the unfolding process.}

\smallskip
A predicate {\em has basic types} if {\em all\/} its arguments have basic type.
An atom has basic types if its predicate has basic types. 
A clause  has basic types if {\em all\/} its atoms have basic types. 

\begin{definition}[Sharing Blocks]\label{def:sharing}
 Let $S$ be a set (or a conjunction) of atoms. 
 For any two atoms $A_1$ and $A_2$ in $S$, 
$A_1 \downarrow A_2$ holds 
iff $\mathit{nbvars}(A_1) \cap \mathit{nbvars}(A_2) \neq \emptyset$.
Let~$\Downarrow$ be the reflexive transitive closure of $\downarrow$.
By $\mathit{SharingBlocks}(S)$ we denote the partition of $S$ into subsets, 
called {\em sharing blocks},
with respect to~$\Downarrow$.
\end{definition}


Algorithm~$\mathcal E$ makes use of the
well-known transformation rules
{\em define}, {\em fold}, {\em unfold}, and {\em replace} for 
CHCs~\cite{EtG96,TaS84}.

\noindent \hrulefill\nopagebreak

\noindent {\bf The Elimination Algorithm}~$\mathcal E$.\\
{\em Input}: A set $\mathit{Cls} \cup \mathit{Gs}$, where $\mathit{Cls}$ is a set of definite clauses and $\mathit{Gs}$ 
is a set of goals;
\\
{\em Output}: A set $\mathit{TransfCls}$ of clauses such that: 
(1) $\mathit{Cls}\cup \mathit{Gs}$ is satisfiable iff $\mathit{TransfCls}$ is satisfiable, and
(2) every clause in $\mathit{TransfCls}$ has basic types.

\vspace*{-2mm}
\noindent \rule{2.0cm}{0.2mm}

\noindent $\mathit{Defs}:=\emptyset$;
\noindent $\mathit{InCls}:=\mathit{Gs}$;
\noindent $\mathit{TransfCls}:=\emptyset;$

\noindent
{\bf while} $\mathit{InCls}\!\neq\!\emptyset$ {\bf do}

$\mathit{Define\mbox{-}Fold}(\mathit{Defs},\mathit{InCls},
\mathit{NewDefs},\mathit{FldCls});$

$\mathit{Unfold}(\mathit{NewDefs},\mathit{Cls},\mathit{UnfCls});$

$\mathit{Replace}(\mathit{UnfCls}, \mathit{Cls}, \mathit{RCls});$

$\mathit{Defs}:=\mathit{Defs}\cup\mathit{NewDefs};$~~
$\mathit{InCls}:=\mathit{RCls};$~~
$\mathit{TransfCls}:=\mathit{TransfCls}\cup\mathit{FldCls};$\nopagebreak

\vspace*{-2mm} 
\noindent \hrulefill


\smallskip
\noindent
Starting from a set $\mathit{Cls}$ of definite clauses and a set $\mathit{Gs}$ of goals,
Algorithm  $\mathcal E$ applies iteratively the procedures $\mathit{Define\mbox{-}Fold}$,
$\mathit{Unfold}$, and $\mathit{Replace}$, in this order, until it derives a set 
$\mathit{TransfCls}$ of clauses 
whose predicates have basic types only.
Algorithm  $\mathcal E$ collects in $\mathit{Defs}$  the clauses, called {\it definitions}, 
introduced by the {\it Define} steps, and collects in
$\mathit{InCls}$ the clauses to be transformed.
The set $\mathit{Defs}$ of definitions is initialized to the empty set and the set $\mathit{InCls}$ 
of the input goals is initialized to the set {\it Gs} of goals.

Now let us present the various procedures used by Algorithm~$\mathcal E$.


	
\noindent \hrulefill \nopagebreak

\noindent {\bf Procedure $\mathit{Define\mbox{-}Fold}(\mathit{Defs}, \mathit{InCls},
	\mathit{NewDefs},\mathit{FldCls})$}
\\
{\em Input}\/: A set {\it Defs} of definitions and a set {\it InCls} of clauses;
\\
{\em Output}\/: A set {\it NewDefs} of definitions and a set $\mathit{FldCls}$ of clauses.
\nopagebreak

\vspace{-2.5mm} \noindent \rule{2.0cm}{0.2mm}


\noindent $\mathit{NewDefs} := \emptyset; \ \mathit{FldCls}:= \emptyset$;

\noindent {\bf for} each clause $C$: $H\leftarrow c, B$ in $\mathit{InCls}$ {\bf do}

\noindent 
\hspace{5mm}
{\bf if} $C$ is a constrained fact {\bf then}
$\mathit{FldCls}:=\mathit{FldCls}\cup\{C\}$ {\bf else}

\noindent 
\hangindent=14mm
\hspace{8mm}
{\underline{\it Define.}} Let $\mathit{SharingBlocks}(B) = \{B_1, \ldots, B_n\}$;
\\
{\bf for} $i=1,\ldots,n$ {\bf do}
\\
{\bf if} there is no clause in  $\mathit{Defs}\cup
\mathit{NewDefs}$ whose body is $B_i$ (modulo the names of variables and the order and
multiplicity of the atoms) {\bf then}
\\
$\mathit{NewDefs} := \mathit{NewDefs} \cup \{\mathit{newp_i}(V_{i})\leftarrow B_i\}$
\\
where:
(i) $\mathit{newp}_i$ is a new predicate symbol, and
(ii)~$V_{i}$ is the tuple of distinct variables of basic type occurring in $B_i$;

\noindent  
\hangindent=14mm
\hspace{8mm}
{\underline{\it Fold.\raisebox{-.7mm}{\rule{0mm}{2mm}}} $C$ is folded using the definitions
	in $\mathit{Defs}\cup \mathit{NewDefs}$, thereby deriving 
	\\
	$F$: $H\leftarrow c,
	\mathit{newp}_1(V_{1}),\ldots, \mathit{newp}_n(V_{n})$
	\\
	where, for $i=1,\ldots,n$, 
	$\mathit{newp}_i(V_{i}) \leftarrow B_i$ is the unique clause in 
	$\mathit{Defs}\cup \mathit{NewDefs}$ whose body is $B_i$, 
	modulo variable renaming;
	\\
	\noindent $\mathit{FldCls}:=\mathit{FldCls}\cup\{F\}$;

	
	\nopagebreak \vspace*{-2mm} \noindent \hrulefill
	
	\smallskip

The $\mathit{Define\mbox{-}Fold}$ procedure removes the arguments with non-basic types from
each clause~$C$ of $\mathit{InCls}$ in two steps: first, the {\it Define} step 
introduces a new predicate definition for each sharing block 
of the body of $C$ 
{(unless a definition with a body equal, modulo variable renaming, to that block 
has already been introduced
in a previous {\it Define} step),} and then, the {\it Fold} step replaces
each sharing block by the head of the new definition. 
Since the heads of these new definitions
have basic types, the body of the clause of the form:
$H\leftarrow c, \mathit{newp}_1(V_{1}),\ldots,$ $ \mathit{newp}_n(V_{n})$,
derived from $C$ by the {\it Fold} step, has basic types.
Also $H$ has basic types, because:
(i)~Algorithm~$\mathcal E$ initializes $\mathit{InCls}$ to the set {\it Gs} of goals, 
whose heads are {\it false},
and (ii)~the head predicate of each new clause added to $\mathit{InCls}$
has been introduced by a previous {\it Define} step,
and hence, by construction, has basic types.

\begin{example}$({\it Tree\su Processing})$. \label{ex:TreeP1}\nopagebreak
Let us consider the introductory example of Section~\ref{sec:Intro-Ex}.
At the first iteration of the while-do body of Algorithm~$\mathcal{E}$,
$\mathit{InCls}$ consists of clause~8,
which is the result of translating the property $\mathit{Prop}$  to be verified. The {\it Define} step introduces
the new predicate $\mathit{new}1$ through the following definition:

\smallskip

\noindent
15.  $\mathit{new}1(N,M,K)\If \leftdrop(N,T,U),\ \minleaf\,(U,M),\ \minleaf\,(T,K)$

\smallskip
\noindent
The body of this clause consists of the single sharing block of the body of clause~8 (note that $\leftdrop(N,T,U)$
shares the tree variables {\it T} and {\it U} with $\minleaf\,(T,K)$ and
$\minleaf\,(U,M)$, respectively).
Now, the {\it Fold} step derives {clause~14} of Section~\ref{sec:Intro-Ex},
where all predicates have arguments of integer type, and thus it is added to the final set {\it TransfCls}.
\end{example}


The {\it Define} step adds to {\it Defs} a set {\it NewDefs} of new definitions, 
whose body may have predicates with non-basic types (see clause 15).
Now, Algorithm~$\mathcal E$
proceeds by applying the {\it Unfold} procedure to those clauses.


\noindent \hrulefill

\noindent {\bf Procedure $\mathit{Unfold}(\mathit{NewDefs},\mathit{Cls},\mathit{UnfCls})$}
\\
{\em Input}\/: A set $\mathit{NewDefs}$ of definitions and a set $\mathit{Cls}$ of definite clauses;
\\
{\em Output}\/: A set $\mathit{UnfCls}$ of clauses.

\vspace*{-2.5mm} \noindent \rule{2.0cm}{0.2mm}


\noindent
$\mathit{UnfCls} := \mathit{NewDefs}$;

\noindent Initially, all atoms in the body of the clauses of $\mathit{UnfCls}$ are marked  `unfoldable';

\noindent 
{\bf while} there exists a clause $C$ in $\mathit{UnfCls}$ of the form
\mbox{$H \leftarrow c, {L}, A, {R}$} such that
either
\\
(i)~the atom $A$ is `unfoldable' and is strictly maximal in ${L}, A, {R}$, or
\\
(ii)~all atoms in ${L}, A, {R}$ are `unfoldable' and  not strictly maximal {\bf do}

\noindent\hspace{5mm}Let $K_{1}\leftarrow c_{1},
B_{1},~\ldots,~K_{m}\leftarrow c_{m}, B_{m}$ be all clauses \hangindent=5mm 
of $\mathit{Cls}$ (where, without loss of generality, we assume $\vars(\mathit{Cls})\cap\vars(C)=\emptyset$) 
such that, for \( i\eq 1,\ldots ,m
\), (1)~there exists a most general unifier $\vartheta_i$ of $A$ and
$K_i$, and (2)~the constraint $(c, c_{i})\vartheta_i$ is
satisfiable;

\smallskip

\noindent
\hspace{5mm}$\mathit{UnfCls}:=(\mathit{UnfCls}-\{C\})\cup \{(H\leftarrow c, c_{1},
{L}, B_{1}, {R})\vartheta_1,
\ldots,
(H\leftarrow c, c_{m}, {L}, B_{m},
{R})\vartheta_m\}$

\smallskip

\noindent
\hspace{5mm}\hangindent=5mm where, for $i=1,\ldots,m$, 
(1)~an atom $E\vartheta_i$ of $(L,R)\vartheta_i$ is `unfoldable' iff 
the corresponding atom $E$
of $(L,R)$ is `unfoldable' in $C$, and
(2)~no atom in $B\vartheta_i$ is `unfoldable';

\nopagebreak \vspace*{-1mm} \noindent \hrulefill

\medskip
The {\it Unfold} procedure unfolds the atoms occurring in 
$\mathit{NewDefs}$ by performing resolution steps with clauses in {\it Cls}.
The procedure applies a strategy that consists in
unfolding strictly maximal atoms, if any.
In the case where the predicates 
are defined by induction on the structure of their arguments with 
non-basic types, this strategy 
corresponds to a form of induction on the arguments structure.
The subsequent folding steps correspond to applications of the 
inductive hypotheses.
A characterization of a class of CHCs where this strategy is successful
will be given in Section~\ref{sec:Correctness}.
The use of the `unfoldable' marking on atoms enforces a finite number of resolution steps.

{Note that when Case~(ii) of the
	condition of the while-do holds, 
	the {\it Unfold} procedure may unfold {\em any\/} `unfoldable' atom 
	(and in our implementation of the procedure we unfold the leftmost one).
	However, in the class of CHCs presented in Section~\ref{sec:Correctness},
	the termination of Algorithm~$\mathcal E$ is independent of the choice of the
	atom to be unfolded.}

\begin{example} $(${\it Tree\su Processing}, {\it continued\/}$).$ Let us continue the 
	derivation presented in Example~\ref{ex:TreeP1}.
All atoms in the body of clause 15 are `unfoldable' and none
is strictly maximal (indeed, no tree argument is a strict superterm of another).
In this case the {\it Unfold} procedure may unfold any `unfoldable'
atom, and
we assume that it unfolds
$\leftdrop(N,T,U)$. 
The following three clauses,
where we have underlined the `unfoldable' atoms, are derived:


\smallskip
\noindent 
16.  $\mathit{new}1(N,M,K)\If \underline{\minleaf\,(\mathit{leaf},M)},\ 
\underline{\minleaf\,(\mathit{leaf},K)}$

\noindent 
17.  $\mathit{new}1(N,M,K)\If N\lesseq 0,\ \underline{\minleaf\,(\mathit{node}(X,L,R),M)},\  
\underline{\minleaf\,(\mathit{node}(X,L,R),K)}$ 


\noindent 
18.  $\mathit{new}1(N,M,K)\If N\greq 1,\  N1 \eq N \minus 1,\  \leftdrop(N1,L,U), \  
\underline{\minleaf\,(U,M)},$

\vspace*{-.4mm}
\hspace*{15mm}$\underline{\minleaf\,(\mathit{node}(X,L,R),K)}$

\smallskip
\noindent
Now the {\it Unfold} procedure continues by selecting more atoms for unfolding.
For instance, in clause~18 it selects $\minleaf\,(\mathit{node}(X,L,R),K)$,
which is strictly maximal because the argument $L$ of $\leftdrop(N1,L,U)$
is a strict subterm of $\mathit{node}(X,L,R)$ and no atom in the body
of clause~18 has an argument that is a strict superterm of $\mathit{node}(X,L,R)$.
{After~five~iterations, where all underlined atoms, except  $\minleaf\,(U,M)$, 
are unfolded, the {\it Unfold} procedure derives the following clauses:}

	
\smallskip
\noindent 
19.  $\mathit{new}1(N,M,K)\If M\eq 0,\ K\eq 0$

\noindent 
20.  $\mathit{new}1(N,M,K)\If N\lesseq 0,\ M\eq M3\plus 1,\ K\eq K3\plus 1,\ 
\minleaf\,(L,M1),\ \minleaf\,(R,M2), $\nopagebreak

\hspace*{15mm}$\mathit{min}(M1,M2,M3),\ \minleaf\,(L,K1),\ \minleaf\,(R,K2),\ \mathit{min}(K1,K2,K3)$

\noindent 
21.  $\mathit{new}1(N,M,K)\If N\greq 1,\ N1 \eq N \minus 1\ K\eq K3\plus 1,\ \leftdrop(N1,L,U),\ \underline{\minleaf\,(U,M)},$\nopagebreak

\hspace*{15mm}$\minleaf\,(L,K1),\ \minleaf\,(R,K2),\ \mathit{min}(K1,K2,K3)$

\smallskip

\noindent 
In clause 21, $\minleaf\,(U,M)$ is an `unfoldable' atom, but the {\it Unfold} procedure
stops because that atom is not strictly maximal.
Clause  19 is a constrained fact and it is added to the final set
{\it TransfCls} (indeed, clause 19 is clause~9 of Section~\ref{sec:Intro-Ex}).
\end{example}

After the {\it Unfold} procedure Algorithm~$\mathcal E$
may simplify some clauses
by exploiting functional dependencies among
predicate arguments, which hold by construction 
for the predicates obtained by translating functional programs.

\begin{definition}[Predicate Functionality]\label{def:fun}
Let $\mathit{Cls}$ be a set of definite clauses.
A predicate $p(X,Y)$, where $X$ and $Y$ are tuples of arguments,
is {\em functional} in $\mathit{Cls}$ if 
$\mathit{Cls} \cup \{ \false \If Y\diff Z,\ p(X,Y),\ p(X,Z)\}$
is a satisfiable set of clauses.
\end{definition}


\noindent \hrulefill \nopagebreak

\noindent {\bf Procedure $\mathit{Replace}(\mathit{UnfCls},\mathit{Cls},\mathit{RCls})$}\nopagebreak
\\
{\em Input}\/: A set $\mathit{UnfCls}$ of clauses and a set $\mathit{Cls}$ of definite clauses;
\\
{\em Output}\/: A set $\mathit{RCls}$ of clauses.\nopagebreak

\vspace{-2.5mm} \noindent \rule{2.0cm}{0.2mm}


\noindent
$\mathit{RCls} := \mathit{UnfCls}$;\nopagebreak

\noindent
{\bf while} there is a clause $C\!\in\!\mathit{RCls}$ of the form:
\mbox{$H \leftarrow c, G_1,\, p(t,u),\, G_{2},\, p(t,w),\, G_{3}$,} 
where predicate $p(X,Y)$ is functional in $\mathit{Cls}$
{\bf do}

\hangindent=5mm Replace $C$ by 
\mbox{$(H \leftarrow c, G_1, p(t,u), G_{2}, G_{3}) \vartheta$,} 
where $\vartheta$ is {a most general unifier} of $u$ and~$w$;


\vspace*{-2mm} \noindent \hrulefill

 
\begin{example}$(${\it Tree\su Processing}, {\it continued\/}$)$.
By the functionality of the predicates $\minleaf\,(T,M)$ and $\mathit{min}(X,Y,Z)$,
clause~20 is replaced by the following one:

\smallskip

\noindent 
22.  $\mathit{new}1(\!N,\!M,\!K)\!\If\! N\!\!\lesseq 0, M\!\!\eq\! M3\plus 1, 
K\!\!\eq\!M,\ \minleaf(L,\!M1),\ \minleaf(R,\!M2),$ 

\hspace*{29mm}$\mathit{min}(M1,\!M2,\!M3)$ 

\smallskip

\noindent 
Now, Algorithm $\mathcal{E}$ performs a second iteration 
by executing again the {\it Define}-{\it Fold},
{\it Unfold}, and {\it Replace} procedures. The {\it Define} step introduces the
following definition:

\smallskip

\noindent 
23.  $\mathit{new}2(M1) \If \minleaf\,(L,M1)$

\smallskip
\noindent
whose body consists of a sharing block in the body of clause~22 (note that the tree variable $L$
is not shared with any other atom). No other definitions are introduced, as all other
sharing blocks in the clauses currently in {\it InCls} (namely, clauses~19, 21, and 22)
are variants of the body of the definitions 15 and 23. 
The {\it Fold} step derives clauses~10 and~11 of Section~\ref{sec:Intro-Ex} from clauses~22 and~21, respectively.

{Finally, Algorithm $\mathcal{E}$ performs an {\it Unfold} step followed by a {\it Fold} step on clause~23
and derives clauses~12 and~13 of Section~\ref{sec:Intro-Ex}. Note that {\it Replace} 
steps are not applicable, and no new definitions are introduced by the {\it Define} step.}
Thus, Algorithm $\mathcal E$ terminates
and returns clauses~1,\,2,\,9--14 of Section~\ref{sec:Intro-Ex}.
\end{example}

	\section{Correctness and Termination of the Transformation}
	\label{sec:Correctness}
	The partial correctness of Algorithm~$\mathcal E$ follows from 
well-known satisfiability preservation results that hold for 
the fold/unfold transformation rules~\cite{EtG96,TaS84}
(see the Appendix for a proof).

\begin{theorem}[Partial Correctness]\label{thm:sound}
	Let $\mathit{Cls}$ be a set of definite clauses and let $\mathit{Gs}$ be a set of goals.
	If Algorithm $\mathcal E$ terminates for the input clauses 
	$\mathit{Cls}\cup \mathit{Gs}$, returning a set $\mathit{TransfCls}$ of clauses,
	then (1)~$\mathit{Cls}\cup \mathit{Gs}$ is satisfiable iff $\mathit{TransfCls}$ is satisfiable, and
	(2)~all clauses in $\mathit{TransfCls}$ have basic types.
\end{theorem}


Now we introduce a class of CHCs where Algorithm~$\mathcal E$
is guaranteed to terminate. 

{First we need the following terminology and notation. An atom\,is\,said\,to\,be\,{\em linear} if 
each\,variable\,occurs\,in\,it\,at\,most\,once.}\,By\,$\mathit{arity}(p)$\,we 
denote\,the\,arity\,of\,predicate symbol $p$. 

\begin{definition}[Slice Decomposition]
Let $C$ be a clause of the form $A_0 \If c, A_1,\ldots, A_m$,
and let ${\mathit{Pred}}$ be the set of predicate symbols in the atoms
$A_0,A_1,\ldots, A_m$ of $C$. A {\em slice} of $C$ is a total function $\sigma: {\mathit{Pred}} \rightarrow {\mathbb N}$
such that, for every predicate $p\!\in \!{\mathit{Pred}}$: 
$(1)$~$0\lesseq\sigma(p) \lesseq \mathit{arity}(p)$, and 
$(2)$~if $\sigma(p)\eq i\gr 0$, then the $i$-th argument of~$p$ has a non-basic type.
For any atom $p(t_1,\ldots,t_k)$ in $C$, we define:

\smallskip

%

$\sigma(p(t_1,\ldots,t_k))=\left\{
\begin{array}{ll}
	p_0 &\mbox{if } \sigma(p)\eq 0\\
	 p_i(t_i) &\mbox{if } \sigma(p)\eq i \!>\!0
\end{array}
\right.$

%

\smallskip

\noindent
where $p_0$ (of arity 0) and $p_i$ (of arity 1) are fresh, new predicate symbols. %


We also define
$\sigma(C)$ to be the new clause $\sigma(A_0) \If \sigma(A_1),\ldots, \sigma(A_m)$.
The slice $\sigma$ of~$C$ is {\em quasi-descending} if 
$\sigma(A_0), \sigma(A_1), \ldots, \sigma(A_m)$ are linear atoms and, for $i=1,\ldots,m$,
$(1)$~for $j=1,\ldots,m$, $\vars(\sigma(A_i)) \cap \vars(\sigma(A_j))\eq \emptyset$, if $i\diff j$, and 
$(2)$~either $\vars(\sigma(A_0)) \cap \vars(\sigma(A_i))\eq \emptyset$ or $\sigma(A_0) \succeq \sigma(A_i)$.

A {\em slice decomposition} of clause~$C$ is a set $\Sigma_{C} = \{\sigma_1,\ldots,\sigma_n\}$, where
$\sigma_1,\ldots,\sigma_n$ are slices of~$C$ and, for all $p\!\in\! \mathit{Pred}$, for all $i\!\in\! \{1,\ldots,\mathit{arity}(p)\}$,
if the $i$-th argument of $p$ has non-basic type, then there exists $\sigma_j\!\in\! \Sigma_{C}$ such that $\sigma_j (p)\!=\!i$. 
%
A~slice decomposition 
$\Sigma_{C}$ of a clause~$C$
is said to be {\em disjoint} if, for any two slices $\sigma_h$ and $\sigma_k$ in $\Sigma_{C}$, we
have that $\vars(\sigma_h(C)) \cap  \vars(\sigma_k(C)) \eq \emptyset$, whenever $h\diff k$.
$\Sigma_{C}$ is said to be {\em quasi-descending} if all slices in $\Sigma_{C}$ are quasi-descending.
\end{definition}

The following are slices of clauses~4 and~7 (see Section~\ref{sec:Intro-Ex}):

\smallskip

$\sigma_{1}(\mathrm{clause}~4) = \mathit{\minleaf}_{1}(\mathit{node}(X,\!L,\!R))\If \mathit{\minleaf}_{\,1}(L), \mathit{\minleaf}_{1}(R), \mathit{min}_{0}$

$\sigma_{2}(\mathrm{clause}~7) = \mathit{\leftdrop}_{2}(\mathit{node}(X,L,R))\If
\mathit{\leftdrop}_{2}(L)$

$\sigma_{3}(\mathrm{clause}~7) = \mathit{\leftdrop}_{3}(T)\If \mathit{\leftdrop}_{3}(T)$

\smallskip
\noindent
We have that $\Sigma_4=\{\sigma_1\}$ 
is a disjoint, quasi-descending slice decomposition of clause~4, and so is $\Sigma_7=\{\sigma_2,\sigma_3\}$
for clause~7.

Now we define a class of goals for which Algorithm~$\mathcal E$ terminates.

\begin{definition}[Sharing Cycle]
	Let $G$ be a goal of the form $\false \If  c,A_1,\ldots,A_m$.
	We say that $G$ has a {\em sharing cycle} if there is a sequence of {$k\,(>\!1)$}
	distinct variables $X_0, \ldots, X_{k-1}$ of non-basic type, and a sequence
	$A'_0,\ldots, A'_k$ of atoms in $A_1,\ldots,A_m$,  such that: 
	(i)~$A'_0\eq A'_k$, 
	(ii)~$A'_1,\ldots, A'_k$ are all distinct, and
	(iii)~for $i=0,\ldots,k\minus 1$, $A'_{i}$ and $A'_{i+ 1}$ share the non-basic variable~$X_i$.
\end{definition}

Now, if we represent the body of goal~8 of Section~\ref{sec:Intro-Ex} as the following labeled graph:

\smallskip
		\begin{tikzpicture}[node distance=1cm, thick]
		\node (ld) at (4.5,0) {{\leftdrop$(N,T,U)$}};
		\node (ml1) at (0,0) {{\minleaf$(T,K)$}};
		\node (ml2) at (9,0) {{\minleaf$(U,M)$}};	
		\draw [-,thick] (ld) -- (ml1);
		\draw [-,thick] (ld) -- (ml2);	
		\node (T) at (2.2,.21) {{$T$}};
		\node (U) at (6.9,.21) {{$U$}};
		\end{tikzpicture}

\noindent 
where the arc from a node $A$ to a node $B$ has label $X$ iff $X \!\in\! 
{\mathit{nbvars}}({A}) \cap {\mathit{nbvars}}({B})$, then it is easy to see that goal~8 has no sharing cycle (indeed, there are no cycles in the above graph).

We have the following result, whose proof is given in the Appendix.


\begin{theorem}[Termination]\label{thm:termination}
Let $\mathit{Cls}$  be a set of definite clauses such that every clause in $\mathit{Cls}$
has a disjoint, quasi-descending slice decomposition.
Let $\mathit{Gs}$ be a set of goals such that, for each goal $G\in\mathit{Gs}$, 
(i) $G$ is of the form
$\false \If  c,A_1,\ldots,A_m$, where for $i\!=\!1,\ldots,m$,
$A_i$ is an atom whose arguments are distinct variables, and
(ii) $G$ has no sharing cycles.
Then Algorithm $\mathcal E$ terminates for the input clauses $\mathit{Cls}\cup\mathit{Gs}$.
\end{theorem}

Note that all definite clauses (clauses 1--7) of our introductory example have a 
quasi-descending slice decomposition (see above for  clauses 4 and 7).
These slice decompositions are also disjoint, except for the one of clause~6
 where the second and third argument of $\leftdrop$ share some variables.
However, clause~6 can be rewritten as:

\smallskip

$\mathit{\leftdrop}(N,\mathit{node}(X,L,R),\mathit{node}(X1,L1,R1))\If N\lesseq 0,\ X\eq X1,\ eqt(L,L1),\ eqt(R,R1)$

$eqt(\mathit{leaf},\mathit{leaf})\If$

$eqt(\mathit{node}(X1,L1,R1),\mathit{node}(X2,L2,R2))\If X1\eq X2,\ eqt(L1,L2),\ eqt(R1,R2)$

\smallskip

\noindent
where predicate $eqt$ defines the equality between binary trees. 
These three clauses have a disjoint,
quasi-descending slice decomposition, and after this rewriting the termination of 
Algorithm~$\mathcal E$ is guaranteed.
%
The rewriting of any constrained fact, such as clause~6, into a clause that
	has a disjoint, quasi-descending slice decomposition 	
	can be done automatically as a pre-processing step, 
	by introducing an equality predicate for each non-basic type. 
	However, in the benchmark set presented in Section~\ref{sec:Experiments}, 
	this pre-processing step has no effect on the termination behavior of our 
	transformation algorithms.

    \section{Adding Integer and Boolean Constraints} 
	\label{sec:Generalization}
	Algorithm $\mathcal E$ is not guaranteed to terminate
outside the class of definite clauses and goals considered in Theorem~\ref{thm:termination}.
Let us consider, for instance, the following set of clauses which specifies a verification problem
on lists: 

\medskip
\noindent
\makebox[64mm][l]{$\mathit{append}([~],\mathit{Ys},\mathit{Ys}) \If$}
\makebox[68mm][l]{$\mathit{take}(N,[~],[~]) \If$}\nopagebreak
  
\noindent
\makebox[64mm][l]{$\mathit{append}([X|\mathit{Xs}],\mathit{Ys},[Z|\mathit{Zs}]) \If X\eq Z,\ $}
\makebox[68mm][l]{$\mathit{take}(N,[X|\mathit{Xs}],[~]) \If N \eq 0$}\nopagebreak

\noindent
\makebox[6mm][l]{}\makebox[58mm][l]{$\mathit{append}(\mathit{Xs},\mathit{Ys},\mathit{Zs})$}
\makebox[68mm][l]{$\mathit{take}(N,[X|\mathit{Xs}],[Y|\mathit{Ys}]) \If
  N \diff 0,\ X\eq Y,\
 $}\nopagebreak
  
\noindent
\makebox[64mm][l]{}
\makebox[6mm][l]{}\makebox[60mm][l]{$ N1 \eq N\minus 1,\ \mathit{take}(N1,\mathit{Xs},\mathit{Ys})$}

\noindent
\makebox[64mm][l]{$\mathit{drop}(N,[~],[~]) \If$}
\makebox[68mm][l]{$\difflist([~],[Y|\mathit{Ys}])\If$}

\noindent
\makebox[64mm][l]{$\mathit{drop}(N,[X|\mathit{Xs}],[Y|\mathit{Xs}]) \If N\eq 0, X\eq Y$}
\makebox[68mm][l]{$\difflist([X|\mathit{Xs}],[~])\If$}

\noindent
\makebox[64mm][l]{$\mathit{drop}(N,[X|\mathit{Xs}],\mathit{Ys}) \If N \diff 0, N1 \eq N\minus 1,$}
\makebox[68mm][l]{$\difflist([X|\mathit{Xs}],[Y|\mathit{Ys}]) \If X\diff Y$}

\noindent
\makebox[6mm][l]{}\makebox[58mm][l]{$\mathit{drop}(N1,\mathit{Xs},\mathit{Ys})$}
\makebox[68mm][l]{$\difflist([X|\mathit{Xs}],[Y|\mathit{Ys}]) \If X\eq Y,\ $}

\noindent
\makebox[64mm][l]{}
\makebox[6mm][l]{}\makebox[60mm][l]{$\difflist(\mathit{Xs},\mathit{Ys})$}
  
\smallskip

\noindent
$\false \If 
    M\eq N,\ \mathit{take}(M,\mathit{Xs},\mathit{Ys}),
    \mathit{drop}(N,\mathit{Xs},\mathit{Zs}),
    \mathit{append}(\mathit{Ys},\mathit{Zs},A), \difflist(A , \mathit{Xs})$
    
   
\medskip
\noindent
In these clauses: 
(i)~$\mathit{take}(M,\mathit{Xs},\mathit{Ys})$ holds if 
the list $\mathit{Ys}$ is the prefix of the list $\mathit{Xs}$ up to its $M$-th element,
(ii)~$\mathit{drop}(N,\mathit{Xs},\mathit{Zs})$ holds if list $\mathit{Zs}$ is a suffix of
$\mathit{Xs}$ starting from its $(N \plus 1)$-th element, 
(iii)~$\mathit{append}(\mathit{Ys},\mathit{Zs},A)$ holds if $A$ is the concatenation of the lists
$\mathit{Ys}$ and $\mathit{Zs}$, and
(iv)~$\difflist(A,\mathit{Xs})$ holds if $A$ and $\mathit{Xs}$ are different lists.

The definite clauses listed above satisfy the hypothesis of Theorem~\ref{thm:termination} 
(after rewriting some constrained facts as done at the end of the previous section), but 
the goal does not. Indeed, that goal has the following sharing cycle:

\newcommand{\linearc}{\raisebox{.4mm}{\rule{6mm}{0.2mm}}}

\smallskip

\noindent
\hspace*{2mm}$\mathit{take}(M,\!\mathit{Xs},\!\mathit{Ys})$ $\stackrel{\displaystyle\mathit{X\!s}}{\linearc\!\!\linearc}$
$\mathit{drop}(N,\!\mathit{Xs},\!\mathit{Zs})$ $\stackrel{\displaystyle\mathit{Z\!s}}{\linearc\!\!\linearc}$
$\mathit{append}(\mathit{Ys},\!\mathit{Zs},\!A)$ $\stackrel{\displaystyle\mathit{Y\!s}}{\linearc\!\!\linearc}$
$\mathit{take}(M,\!\mathit{Xs},\!\mathit{Ys})$


\smallskip

\noindent
Algorithm $\mathcal E$ does not terminate on this example.
Indeed, starting from the definition:

\smallskip

\noindent
\hspace*{2mm}$\mathit{new}0(M,N) \If 
\mathit{take}(M,\mathit{Xs},\mathit{Ys}),
\mathit{drop}(N,\mathit{Xs},\mathit{Zs}),
\mathit{append}(\mathit{Ys},\mathit{Zs},A),$ $\difflist(A , \mathit{Xs})$

\smallskip
\noindent
infinitely many new predicates with unbounded lists are generated by Algorithm~$\mathcal E$.
However, these predicates correspond to cases where $M\!\diff\! N$, and if we keep
the constraint $M\!\eq\! N$ between the first 
arguments of $\mathit{take}$ and $\mathit{drop}$,
then a finite set of new predicates is generated. In particular, if we start from the definition:

\smallskip

\noindent
\hspace*{2mm}$\mathit{new}1(M,N) \If M\!\eq\! N,
\mathit{take}(\!M,\mathit{Xs},\mathit{Ys}),
\mathit{drop}(\!N,\mathit{Xs},\mathit{Zs}),
\mathit{append}(\!\mathit{Ys},\mathit{Zs},A), 
\difflist(\!A,\mathit{Xs})$

\smallskip

\noindent
the transformation terminates after a few steps and derives the following equisatisfiable
set of clauses, where an equality constraint holds for the two arguments of each occurrence 
of $\mathit{new}1$:

\smallskip

	${\mathit{new}}2 \If {\mathit{new}}2$\nopagebreak
	
	${\mathit{new}}1(M,N) \If M\!\eq\!N,\ M\!\eq\!0,\ {\mathit{new}}2$\nopagebreak
	
	${\mathit{new}}1(M,N) \If M\!\eq\!N,\ M\!\neq\!0,\ M\!\eq\!1\!+\!M1,\ N\!\eq\!1\!+\!N1,\  {\mathit{new}}1(M1,N1)$\nopagebreak
	

	${\mathit{false}} \If M\!\eq\!N,\ {\mathit{new}}1(M,N)$
	

\smallskip

\noindent
The satisfiability of this set of clauses is trivial, because it does not contain any 
constrained fact, and is
 easily proved by the Z3 solver. 

The above example motivates the introduction of a 
variant of Algorithm~$\mathcal E$, called Algorithm~$\EC$, which is  
obtained by allowing in the {\it Define} step 
the introduction of 
definitions whose bodies may include constraints in 
$\mathit{LIA} \cup \mathit{Bool}$, and hence applying {\it Fold} steps
with respect to these definitions.
The rest of Algorithm~$\EC$ is equal to Algorithm~$\mathcal E$.


As usual in constraint-based transformation techniques (see~\cite{De&17a}
for a recent paper), the computation of a suitable constraint when introducing
a new definition is done by means of a {\em constraint generalization} function 
$\mathit{Gen}$.
We say that a constraint~$g$ is {\em more general than}, or
it is a {\em generalization of}, a constraint~$c$, if
$\mathit{LIA}\cup \mathit{Bool} \models \forall ( c \Imp g)$.
Given a constraint $c$, a conjunction $B$ of atoms, and 
a set $\mathit{Defs}$ of definitions, 
the function $\mathit{Gen}$ matches the constraint $c$ against the constraint
$d$ occurring in a definition in $\mathit{Defs}$ whose body is of the form $d, B$ (modulo
variable renaming), and returns a new constraint $\mathit{Gen}(c,B,\mathit{Defs})$
which is more general than both $c$ and $d$.

The details of how the constraint generalization function 
$\mathit{Gen}$ is actually implemented are not necessary for understanding
Algorithm~$\EC$, which is parametric with respect to such a 
function.
Let us only mention here that the function 
$\mathit{Gen}$ used for the experiments reported in Section~\ref{sec:Experiments} makes use
on the {\em widening operator} based on {\em bounded difference shapes},
which is a standard operator on convex polyhedra considered in the 
field of {\em abstract interpretation}~\cite{Bag&08,CoH78}.
The use of widening avoids the introduction of infinitely many 
definitions that differ for the constraints only.

The {\it Define} and {\it Fold} steps for Algorithm~$\EC$ are as follows.

\medskip

\noindent
Let $C$: $H\leftarrow c, B$ in $\mathit{InCls}$ be a clause which is not a constrained fact.

\noindent 
\hspace{5mm}\hangindent=10mm
{\underline{\it Define.}} 
Let $\mathit{SharingBlocks}(B) = \{B_1, \ldots, B_n\}$;
\\[2pt]
{\bf for} $i=1,\ldots,n$ {\bf do} \\
$g_i := \mathit{Gen}(c,B_i,\mathit{Defs}\cup\!
\mathit{NewDefs})$; 
\\
{\bf if} there is no clause in  $\mathit{Defs}\cup\!
\mathit{NewDefs}$ whose body is $g_i, B_i$
\\
{\bf then} $\mathit{NewDefs} := \mathit{NewDefs} \cup \{\mathit{newp_i}(V_{i})\leftarrow g_i, B_i\}$
\\
where:
(i) $\mathit{newp}_i$ is a new predicate symbol, and
(ii)~$V_{i}$ is the tuple of distinct variables of basic type occurring in $B_i$;

\noindent  
\hspace{5mm}\hangindent=10mm
{\underline{\it Fold.\raisebox{-.7mm}{\rule{0mm}{2mm}}} $C$ is folded using the definitions
	in $\mathit{Defs}\cup \mathit{NewDefs}$, thereby deriving 
	\\[2pt]
	$F$: $H\leftarrow c,
	\mathit{newp}_1(V_{1}),\ldots, \mathit{newp}_n(V_{n})$
	\\
	where, for $i=1,\ldots,n$, $\mathit{newp}_i(V_{i}) \leftarrow  g_i, B_i$ is the unique clause in 
	$\mathit{Defs}\cup \mathit{NewDefs}$ whose body is  $g_i, B_i$, 
	modulo variable renaming;\\
	\noindent $\mathit{FldCls}:=\mathit{FldCls}\cup\{F\}$;
	

\smallskip

\noindent
The partial correctness of Algorithm $\mathcal E$ (see Theorem~\ref{thm:sound})
carries over to Algorithm $\EC$, because also $\EC$ can be expressed as
a sequence of applications of the fold/unfold transformation rules
for CHCs~\cite{EtG96}.
Also the termination result for Algorithm $\mathcal E$ (see Theorem~\ref{thm:termination})
extends to $\EC$, as long as the function $\mathit{Gen}$ guarantees
that, for a given conjunction $B$ of atoms, 
finitely many definitions of the form $\mathit{newp}(V)\leftarrow g, B$ can be introduced.


	\section{Experimental Evaluation} 
	\label{sec:Experiments}
	In this section we present the results of an experimental evaluation we have 
performed for assessing the effectiveness of our approach and, in particular,
for comparing it with 
the approach that extends CHC solvers by adding inductive rules, as done in the
\rcaml~tool \cite{Un&17} 
based on the Z3 solver. 

\paragraph{Implementation.} 
We have implemented the transformation strategy presented in Section~\ref{sec:Transform}
using the \verimap system~\cite{De&14b} together with the Parma Polyhedra
Library (PPL)~\cite{Bag&08} for performing constraint generalizations. 
Then, we have used the \zthree solver v4.6.0 with the \spacer fixed-point
engine~\cite{Ko&13} to check the satisfiability of the transformed CHCs.

The tool and the benchmark suite are available at {\small \url{https://fmlab.unich.it/iclp2018/}}.

\paragraph{Benchmark suite and experiments.}
Our benchmark suite is a collection of 105 verification problems, 
each one consisting of an OCaml functional program manipulating 
inductively defined data structures (such as lists or trees)
together with a property to be verified.
Most of the problems (70 out of 105) derive from the benchmark suite of~\rcaml 
(see {\small \url{http://www.cs.tsukuba.ac.jp/~uhiro/}}, Software {\it{RCaml}},
[web demo  (induction)]).
This suite, in turn, includes problems from the suite of 
\mbox{IsaPlanner}~\cite{DiF03} which is a generic framework for proof planning in 
the Isabelle theorem prover.
We divide our benchmark suite into four sets of problems (see Table~\ref{tab:evaluation}):
$(1)$~\textit{FirstOrder}, the set of problems
relative to first-order programs (57 out of~70) in the \rcaml suite,
$(2)$~\textit{HigherOrderInstances}, the set of problems relative to first-order 
programs (13 out of 70)  that
have been obtained by instantiating higher-order programs in the \rcaml suite,
$(3)$~\textit{MoreLists}, a set of 16 verification problems on lists, and
$(4)$~\textit{MoreTrees}, a set of 19 verification problems on trees.
In our benchmark 94 programs do satisfy the associated property and 
the remaining 11 do not. 

By using the preprocessor provided by the \rcaml system,
each verification problem has been translated 
into a set of CHCs (see the translation {\it Tr} of Section~\ref{sec:Prelim}).
Then, for each derived set, call it $I$, of CHCs we have performed the following three experiments,
whose results are summarized in Table~\ref{tab:evaluation}. 
(A table with the 
detailed results for each problem of the benchmark is available at {\small{\url{https://fmlab.unich.it/iclp2018/}}}.)

\smallskip


\noindent\hangindent=4mm
$\bullet$ 
We have run \zthree 
(which does not use any structural induction rule)
for checking the satisfiability of $I$ (see the two columns for \zthree).

\noindent\hangindent=4mm
$\bullet$ 
We have applied the transformation algorithm~$\EC$
to $I$, thereby producing a set~$T$ of CHCs,
and then we have run \zthree for checking the satisfiability of  $T$
(see the two columns for $\EC$;\zthree). 

\noindent\hangindent=4mm
$\bullet$ 
We have run \rcaml on $I$ (see the two columns for \rcaml).

\smallskip

\noindent
For each verification problem 
we have set a timeout limit of 300 seconds.
\zthree and \mbox{\verimap} have run on an Intel Xeon CPU E5-2640 2.00GHz 
with 64GB under CentOS. 
\rcaml has run in a Linux virtual machine on an Intel i5 2.3GHz with 8GB memory under macOS.

\begin{table}[t]
\begin{center}

{\small
\begin{tabular}{l @{\hspace{-6mm}}l r | @{\hspace{-2mm}} r@{\hspace{-1.5mm}} r| r@{\hspace{-1.5mm}} r | r@{\hspace{-1.5mm}} r }
\cline{4-9}\\[-3mm]
\multicolumn{2}{c}{} & 
&
\multicolumn{2}{c|}{ {\zthree}}& 
\multicolumn{2}{c|}{ ${\EC;\zthree}$}&
\multicolumn{2}{c}{ \rcaml}\\[-2.2mm]
\hline\\[-5.4mm]
\multicolumn{2}{c}{Problem Set} & 
$n$ &
~~~~~~$S_\zthree$ &~~~~~~$T_\zthree$ &
~~$S_{\EC;\zthree}$ &~~~~~~$T_{\EC;\zthree}$ &
~~$S_\rcaml$ &~~~~~~$T_\rcaml$ \\[-1.2mm]
\hline\\[-5.4mm]
\textit{$(1)$ FirstOrder}  &	& 57   &    3	& 	0.09		 	&  47 & 37.64 & 41 & 216.59\\[.4mm]
\textit{$(2)$ HigherOrderInstances} &  & 13  & 	1	& 	0.04		  	 &  11 & 8.33  & 10 & 45.40 \\[.4mm]
\textit{$(3)$ MoreLists}  & & 16	  &    3	& 	13.87	 &  14 & 11.27 & 10 & 119.01
\\[.4mm]
\textit{$(4)$ MoreTrees}  &	& 19   &    5 	& 	20.18	 &  19 & 26.79 & 5 & 55.16\\[-1.9mm]
\hline\\[-5.7mm]
~~~~~~~~~~~~~~\textit{Total}  	&	& 105  &   12 	& 	34.18	 &  91 & 84.03 & 66 & 436.17 
\\
~~~~~~~~~~~~~~\textit{Avg time}  &	& 	  &      	& 	2.85  	 &      &  0.92   &       &  6.61 \\[-1mm]
\end{tabular}
}
\caption{{\small
Column $n$ reports the number of problems in each Problem Set.
Columns~$S_\zthree$ and $T_\zthree$ report the number of problems which have been
solved by Z3 and the total time needed for their solution. 
Analogously, for the two columns referring to
 ${\EC;\zthree}$ and the two columns referring to \rcaml.
Times are in seconds. The timeout occurs after 300\secs. 
\label{tab:evaluation}}
}
\end{center}
\end{table}

\paragraph{Results.}
The figures in Table~\ref{tab:evaluation} show  
that our approach considerably increases 
the effectiveness of CHC satisfiability checking.
Indeed, when  directly applied to the CHCs 
that encode the given verification problems,
\zthree  is able to solve {12} problems,
while it solves {91} problems when it is applied to the CHCs produced 
by our transformation algorithm~$\EC$
(compare Columns $S_\zthree$ and $S_{\EC;\zthree}$).
For the remaining {14} problems,
Algorithm~$\EC$ does not terminate within the time limit,
and thus no CHCs are produced.
Note that these 14 problems fall outside 
the class of programs
for which Algorithm~$\mathcal E$, and also Algorithm~$\EC$, is guaranteed
to terminate.

Table~\ref{tab:evaluation} shows that our approach compares quite well with respect to 
the induction based approach implemented in the \rcaml system~\cite{Un&17}. 
Indeed, $\EC;\zthree$ proves {65} out of {66} problems that are also proved by  \rcaml, 
and in addition it proves {26} 
problems, 7 of which belong to the \rcaml benchmark suite.
Also the average times appear to be favorable to $\EC;\zthree$ with respect to \rcaml.
In particular, on the set of 67 problems where both  $\EC;\zthree$ and \rcaml
provide the solution within the timeout, $\EC;\zthree$ is about six times faster than \rcaml,
having taken into account the fact that the machine on which we have run \rcaml  
is approximately $13\%$ slower.

Now let us report on some 
other important facts not shown in Table~\ref{tab:evaluation}.
When solving verification problems via ${\EC;\zthree}$,
a substantial portion of work is performed 
by Algorithm~$\EC$. Indeed, 
the total time spent by \zthree for solving  
the {91} problems is only $6\%$
of the total solving time (84.03\secs). 
Moreover,
for {42} problems the set of clauses produced by~$\EC$ does not contain 
constrained facts, 
and thus its satisfiability can immediately be checked by~\zthree.

Unfortunately, it may happen that our transformation prevents the
solution of some verification problems.
Indeed, the transformation algorithm~$\EC$ does not terminate within the timeout 
on two problems that can be proved using \zthree alone.
However, on the 10 problems that are solved by both \zthree and $\EC;\zthree$,
the average time taken by  $\EC;\zthree$ (3.41\secs) to solve one problem
is much lower than that taken by \zthree alone (9.74\secs).

Finally, let us comment on our benchmark suite.
Many of the problems are small, but the properties to be verified are not trivial,
such as those presented in Sections~\ref{sec:Intro-Ex} and~\ref{sec:Generalization}.
One of the most difficult problems solved by $\EC;\zthree$, but not by
\rcaml, consists in showing that the insertion of a node in a binary search tree
produces again a binary search tree. That solution took about 5\secs.
The solution of some  of the {13} problems
on which  both $\EC;\zthree$ and \rcaml run out of time, 
requires the use of suitable lemmas.
For instance, one of these problems can be proved by using a lemma stating that the sum of 
the elements of the concatenation of two lists is equal to the sum of the elements of the first
list  plus the sum 
of the elements of the second list.
The extension of our approach 
to support automatic lemma discovery is left for future work.

	\section{Related Work and Conclusions}
	\label{sec:RelConcl}
%
We have presented a two-step approach to the verification of programs
that manipulate inductively defined data structures, such as lists and trees.
The first step consists in the transformation of the set of clauses
that encodes the given verification problem,
into a new, equisatisfiable set of clauses 
whose variables do no longer refer to inductive data structures.
The second step consists in the application of a CHC solver 
to the derived set of clauses with integer and boolean constraints only. 
Thus, in our approach we can take full advantage of the many efficient solvers that 
exist for clauses with integer and boolean 
constraints~\cite{DeB08,Gr&12,HoB12,Ho&12,Ko&13}.
Through some experiments we have shown that, using an implementation of our algorithm
and the Z3 solver, our two-step approach is competitive with respect
to other techniques that extend CHC solvers 
by adding deduction rules for reasoning on inductive data structures~\cite{Un&17}. 

Our transformation technique is related to methods proposed in the past 
for improving the efficiency of functional and logic programs, such as {\em deforestation}~\cite{Wad90},
{\em unnecessary variable elimination} (UVE)~\cite{PrP95a}, and {\em conjunctive partial deduction}~\cite{De&99}.
Among these techniques, the UVE transformation, which makes use of the
fold/unfold rules, is the most similar to the one presented in this paper. 
However, in this paper we have introduced several technical novelties with respect to UVE:
(i)~the use of type information, (ii)~the use of constraints, and (iii)~a better characterization of the
termination of the main algorithm (in particular, here we have introduced
the notions of a slice decomposition and a circular sharing). At a more general level, the objective of the work
presented in this paper is to show the usefulness of our transformation techniques for the improvement of
the effectiveness
of CHC solvers, rather than the improvement of the execution of logic programs.

The idea of using a transformation-based approach
for the verification of software comes from the area of Constraint Logic Programming,
where program specialization has been applied as a means of deriving CLP programs from interpreters of
imperative languages~\cite{Al&07,De&17b,Me&07,Pe&98}.
CHC solvers based on combinations of transformations and abstract interpretation have also
been developed~\cite{De&14b,Ka&16} and have been shown to be competitive with
solvers based on CEGAR, Interpolation, and PDR.


Recently, CHCs have been proposed for verifying {\em relational} program properties~\cite{Fe&14},
that is, properties that relate two programs, such as equivalence. 
It has also been shown that {\em predicate pairing},
which is a fold/unfold transformation for CHCs, greatly improves 
the effectiveness of CHC solvers for relational verification~\cite{De&15c,De&16c}.
A related CHC transformation technique, called {\em CHC product}, works
by composing pairs of clauses with an effect similar 
to predicate pairing, although in some cases it may derive sets of clauses
with fewer predicates~\cite{MoF17}.
Neither predicate pairing nor CHC product can remove
inductively defined data structures, as done by the transformation technique presented in this paper.

Similarly to the technique presented in this paper,
fold/unfold transformations and constraint generalization
have also been used in a verification technique for
imperative programs that compute over arrays~\cite{De&17a}. 
However, the above mentioned technique  is not able 
to remove array data structures, and unlike the one presented here,
it does not consider inductively defined data structures.

We plan to extend our transformation-based
verification method in a few directions, and in particular we plan: (i)~to
study the problem of automatically generating the lemmas which
are sometimes needed for removing data structures,
and (ii)~to consider the verification problem for higher-order functional programs.


	\section{Acknowledgements}
	We would like to thank Hiroshi Unno for his support in the use of the \rcaml system.
	We thank the anonymous reviewers for their constructive comments.
	The authors are members of the INdAM Research group GNCS.
	E.~De~Angelis, F.~Fioravanti and A.~Pettorossi are research associates at 
	CNR-IASI, Rome, Italy.

    \newpage
	\section*{Appendix} 
	\label{sec:Appendix}
	In order to prove Theorem~\ref{thm:sound}, we first recall some notions
and results regarding the transformation rules and their correctness.

A {\it transformation sequence} 
is a sequence  
$S_0,S_1,\ldots,S_n$ of sets of CHCs, whose constraints are in $\mathit {LIA} \cup \mathit {Bool}$,
where, for $i\!=\!0,\ldots,n\!-\!1,$ $S_{i+1}$ is derived from $S_i$ by
applying one of the following \mbox{rules~R1--R5}.

Let $\textit{Defs}_i$ denote the set of all the clauses, 
called {\it definitions}, 
introduced by rule~R1 during the construction of the transformation sequence
$S_0,S_1,\ldots,S_i$. 
In particular, $\textit{Defs}_0\!=\!\emptyset$.

\smallskip
\noindent
(R1)~{\it Define.}  
We introduce a clause $D$: $\textit{newp}(X_1,\ldots,X_k)\leftarrow c,G$, 
where:
(i)~\textit{newp} is {a predicate symbol 
not occurring in the sequence $S_0,S_1,\ldots,S_i$,}
(ii)~$c$ is a constraint in ${\mathit{LIA}}\cup{\mathit{Bool}}$,  
(iii)~$G$ is a non-empty conjunction of atoms whose predicate symbols 
occur in $S_0$, and 
(iv)~$X_1,\ldots,X_k$ are distinct variables occurring 
in $(c,G)$.
Then, we derive the new set $S_{i+1}=S_i\cup \{D\}$ and
$\textit{Defs}_{i+1}=\textit{Defs}_i \cup \{D\}$.

\smallskip
\noindent
(R2)~{\it Unfold.} 
Let  $C$: $H\leftarrow c,L,A,R$ be a variant of a clause in $S_i$.
Let $K_{1}\leftarrow c_{1},
B_{1},~\ldots,$ $K_{m}\leftarrow c_{m}, B_{m}$ be all clauses
of $S_i$ (without loss of generality, we assume $\vars(S_i)\cap\vars(C)=\emptyset$) 
such that, for \( j\eq 1,\ldots ,m\), (1)~there exists a most general unifier $\vartheta_j$ of~$A$ and
$K_j$, and (2)~the constraint $(c, c_{j})\vartheta_j$ is
satisfiable.
By {\it unfolding the atom $A$ in~$C$ using $S_i$}
we derive the new set~$S_{i+1}=(S_i\setminus\{C\}) \cup 
\{(H\leftarrow c, c_{1},
{L}, B_{1}, {R})\vartheta_1,
\ldots, (H\leftarrow c, c_{m}, {L}, B_{m}, {R})\vartheta_m\}$.

\smallskip
\noindent
(R3)~{\it Fold.} 
Let $C$: $H\leftarrow c, L,Q,R$ be a clause in $S_i$, where 
$Q$ is a non-empty conjunction of atoms, and let
$D$: $K \leftarrow d, B$ be (a variant of) a clause in $\textit{Defs}_i$
with $\vars(C)\cap \vars(D)\!=\!\emptyset$.
Suppose that there exist a substitution~$\vartheta$ 
and a constraint $e$ such that:
(i)~\mbox{$Q\!=\! B\vartheta $,}
(ii)~${\mathit{LIA}} \cup {\mathit{Bool}} \models \forall(c \leftrightarrow (e\! \wedge\! d\vartheta))$, 
and
(iii)~for every variable \( X\!\in\!\vars(d,B)\setminus\vars(K)\),
the following conditions hold: (iii.1) \(
X\vartheta \) is a variable not occurring in \( \{H,c,L,R\}
\), and (iii.2)~\( X\vartheta  \) does not occur in the term \(
Y\vartheta  \), for any variable \( Y \) occurring in \( (d, B) \)
and different from \( X \).
By \textit{folding \( C\)
	using the definition \( D\)}, we derive clause 
\(E  \):~\( H\leftarrow e, L, K\vartheta, R\).
In this case we also say that $E$ is derived 
\textit{by folding $Q$ in $C$}.
We derive the new set \( S_{i+1}=(S_{i}\setminus\{C\})\cup \{E \} \).

%

\smallskip
\noindent
(R4)~{\it Replace Equivalent Constraints.} 
Let us consider a subset of $S_i$ of the form \linebreak
$\{(H\leftarrow c_1, G), \ldots,(H\leftarrow c_k, G)\}$. 
Suppose that, for some constraints $d_1,\ldots,d_m,$ 

\smallskip

${\mathit{LIA}} \cup {\mathit{Bool}} \models \forall \, (\exists Y_1\ldots\exists Y_r\ (c_1 \vee \ldots \vee c_k) \leftrightarrow 
\exists Z_1\ldots\exists Z_s\ (d_1 \vee \ldots \vee d_m))$

\smallskip

\noindent
where $\{Y_1,\ldots,\!Y_r\}\!=\!\vars(c_1 \vee \ldots \vee c_k)\setminus \vars(\{H,\!G\})$
and $\{Z_1,\ldots,\!Z_s\}\!=\!\vars(d_1 \vee \ldots \vee d_m)\setminus \vars(\{H,G\})$.
Then, we derive the new set 
\noindent
$S_{i+1} = (S_i \setminus\{(H\leftarrow c_1, G), \ldots,$ $(H\leftarrow c_k, G)\})$ 
$ \cup $ $\{(H\leftarrow d_1, G), \ldots,(H\leftarrow d_m, G)\}$.

\smallskip

Note that rule R4 enables the deletion of a clause with an inconsistent constraint 
in its body. Indeed, if $c_1$ is unsatisfiable, then 
${\mathit{LIA}}\cup{\mathit{Bool}} \models \forall \, (c_1 \leftrightarrow d_1 \vee \ldots \vee d_m)$ 
with~$m\!=\!0$.

\smallskip
\noindent
(R5)~{\it Replace Functional Predicates.} 
	Let $C$:
\mbox{$H \leftarrow c, G_1, p(t,u), G_{2}, p(t,w), G_{3}$,} be a clause in $S_i$
and let $p(X,Y)$ be functional in $S_i$ (see Definition~\ref{def:fun}).
Then, we derive the new set
$S_{i+1}= (S_i \setminus \{C\}) \cup \{(H \leftarrow c, G_1, p(t,u), G_{2}, G_{3}) \vartheta\}$, 
where $\vartheta$ is the most general unifier of $u$ and $w$.

\smallskip
The following theorem sums up various results
presented in the literature~\cite{EtG96,TaS84}. 

\begin{samepage}
\begin{theorem}[Equivalence with respect to satisfiability]\label{thm:equiv}\nopagebreak
	Let $S_0,S_1$,\,$\ldots$, $S_n$ be a transformation sequence
	such that every definition in $\textit{Defs}_n$ is unfolded during the 
	construction of this sequence.
	Then, $S_0 \cup {\mathit{LIA}}\cup{\mathit{Bool}}$ is satisfiable if and only if $S_n \cup {\mathit{LIA}}\cup{\mathit{Bool}}$
	is satisfiable.
\end{theorem}
\end{samepage}

Now, we prove Theorems~\ref{thm:sound} and~\ref{thm:termination} of Section~\ref{sec:Correctness}.

\medskip

\noindent
{\it Theorem~\ref{thm:sound} $($Partial Correctness$)$}\\
	Let $\mathit{Cls}$ be a set of definite clauses and let $\mathit{Gs}$ be a set of goals.
	If Algorithm $\mathcal E$ terminates for the input clauses 
	$\mathit{Cls}\cup \mathit{Gs}$, returning a set $\mathit{TransfCls}$ of clauses,
	then (1)~$\mathit{Cls}\cup \mathit{Gs}$ is satisfiable iff $\mathit{TransfCls}$ is satisfiable, and
	(2)~all clauses in $\mathit{TransfCls}$ have basic types.

\noindent
{\it Proof}

\noindent
Point~(1) follows from Theorem~\ref{thm:equiv} by taking into account
that:~(i) Algorithm~$\mathcal E$ can be viewed as a particular sequence of
applications of Rules R1--R5, and (ii)~every definition in $\textit{Defs}$ is unfolded
during the execution of~$\mathcal E$.

\noindent
Point (2) follows from the fact that, by construction,
every clause introduced in $\mathit{TransfCls}$ has basic types.
To see this, let us consider a clause $C$ in $\mathit{TransfCls}$.
Clause~$C$ belongs to the set {\it FldCls} of clauses obtained by a {\it Define}-{\it Fold}
step. Looking at the {\it Define}-{\it Fold} procedure, we have that:
(i)~the head of $C$ is either $\false$ (because $C\!\in\! \mathit{Gs}$)
or its head predicate has been introduced by the {\it Define} step,
and hence, by construction, has basic types, and (ii)~the body of $C$ has the form:
$c, \mathit{newp}_1(V_{1}),\ldots, \mathit{newp}_n(V_{n})$, where 
$c$ is a constraint which has basic types (because it belongs to 
${\mathit{LIA}} \cup {\mathit{Bool}}$) and  
$\mathit{newp}_1,\ldots, \mathit{newp}_n$ are predicates that, by construction,
have basic types. \hfill $\Box$

%
%

\medskip

\noindent
{\it Theorem~\ref{thm:termination} $($Termination$)$}
\\
Let $\mathit{Cls}$  be a set of definite clauses such that every clause in $\mathit{Cls}$
	has a disjoint, quasi-descending slice decomposition.
	Let $\mathit{Gs}$ be a set of goals such that, for each goal $G\in\mathit{Gs}$, 
	(i) $G$ is of the form
	$\false \If  c,A_1,\ldots,A_m$, where for $i\!=\!1,\ldots,m$,
	$A_i$ is an atom whose arguments are distinct variables, and
	(ii) $G$ has no sharing cycles.
	Then Algorithm $\mathcal E$ terminates for the input clauses $\mathit{Cls}\cup\mathit{Gs}$.

\smallskip

\noindent
{\it Proof} (Sketch)

\noindent
Without loss of generality, we assume that for every clause $H\leftarrow c, B$ 
every variable of basic type has at most one occurrence in  $H, B$.

For any tree $t$, by $\mathit{height}(t)$ we denote the {\it height} of $t$, that is,
the maximal length of a 
path from the root of $t$ to one of its leaves. We extend the function $\mathit{height}$
to terms and atoms, viewed as trees.
First of all, we observe that Algorithm $\mathcal E$ terminates iff there exists two
non-negative integers $H$ and $N$ such that, for every definition
$\mathit{newp}(V)\leftarrow A_1,\ldots,A_n$
added to {\it Defs} during the execution of $\mathcal E$, 
we have that, for $i=1,\ldots n,$ $\mathit{height}(A_i)\lesseq H$,
and $n\lesseq N$.
Indeed, the existence of $H$ and $N$ implies the finiteness of
the set of definitions introduced by $\mathcal E$, and hence the finiteness of
the number of iterations of the body of the while-do loop of $\mathcal E$ itself.

Let us consider a clause $D$: $\mathit{newp}(V)\leftarrow A_1,\ldots,A_n$ added to {\it Defs} 
during the execution of $\mathcal E$. 
Then $D$ satisfies the following properties:

\vspace*{-2mm}
\noindent
\begin{itemize}

\item[P1.] The goal $\false \If A_1,\ldots,A_n$ obtained from $D$ by replacing 
$\mathit{newp}(V)$ by $\false$, has no sharing cycles;


\item[P2.] All atoms in the body of $D$ are linear;

\item[P3.] For any two distinct atoms $A_i$ and $A_j$ in the body of $D$, 
if $A_i$ and $A_j$ share a non-basic variable, then they
are of the form $p(\ldots,t_i,\ldots)$ and $q(\ldots,t_j,\ldots)$, 
where either $t_i\preceq t_j$ or $t_j\preceq t_i$, and $A_i$ and $A_j$ 
do not share any variable besides the ones in $\mathit{vars}(t_i) \cap \mathit{vars}(t_j)$;

\item[P4.] For any atom $A_i$ in the body of $D$,
$\mathit{height}(A_i)\lesseq H$, where $H$ is the maximal height of an atom in
${\it Cls}\cup {\it Gs}$;

\item[P5.] Let $V_G$ be the number of occurrences of non-basic variables
in a goal $G$, and let $M$ be $\max \{V_G \mid G \in \mathit{Gs}\} \plus 1$. Then, in the body of $D$,
(P5.1)~every non-basic variable has at most $M$ occurrences, and
(P5.2)~there exist $K\lesseq M$ predicate arguments 
such that every non-basic variable that occurs more than once, 
also occurs in one of those $K$ arguments.
\end{itemize}

\vspace{-2mm}
\noindent
Property~P1 holds for each clause  $D$ initially in {\it Defs} 
by the hypothesis that
$\mathit{Gs}$ is a set of goals that have no sharing cycles.
This property, when referred to the body of the clauses, is preserved by the {\it Unfold} 
and {\it Replace} steps, due to the hypothesis on the clauses in {\it Cls},
and hence it also holds for each new definition added to {\it Defs}
by the {\it Define} step after {\it Unfold} 
and {\it Replace}.


Property~P2 holds for each clause $D$ initially in {\it Defs}
by Hypothesis~(i) on $\mathit{Gs}$.
This property is preserved by the {\it Unfold} 
and {\it Replace} steps, due to the hypothesis on the clauses in {\it Cls}.
Note, in particular, that the existence of a disjoint, quasi-descending slice
decomposition for all clauses in {\it Cls}
implies that each atom in the body of a clause in {\it Cls} is linear, and hence
only linear atoms are introduced by {\it Unfold} steps.
The linearity of the atoms different from the one
replaced by an {\it Unfold} step is enforced by the existence of
a disjoint, quasi-descending slice decomposition for all clauses in {\it Cls}. 
Linearity is also preserved by {\it Replace} steps.
Thus, Property~P2 follows from the fact that the body of $D$ consists of atoms
taken from the body of a clause derived by  {\it Unfold} and {\it Replace} steps.

Property~P3 holds for each clause $D$ initially in {\it Defs}
by Hypothesis~(i) on $\mathit{Gs}$.
Property~P3 also holds for each clause derived by the {\it Unfold} and {\it Replace} steps
by Property~P1 and by the hypothesis that  every clause in $\mathit{Cls}$
has a  disjoint, quasi-descending slice decomposition.
Then, Property~P3 follows from the fact that the body of $D$ consists of atoms
taken from the body of clauses derived by  {\it Unfold} and {\it Replace} steps.

Property~P4 holds for each clause $D$ initially in {\it Defs}
because the body of clause $D$ is the
set of atoms occurring in the body of a goal in {\it Gs}.
This property also holds for each clause derived by {\it Unfold}  steps. 
Indeed, suppose that we unfold an atom $A$ in the clause $C$ of the form
\mbox{$H \leftarrow c, {L}, A, {R}$} such that
either (i)~$A$ is strictly maximal in ${L}, A, {R}$, or
(ii)~all atoms in ${L}, A, {R}$ are  not strictly maximal.
Both in case~(i) and case~(ii), by Property~P3,
$A$ is of the form $p(\ldots,t_i,\ldots)$ and any atom $Q$ in $L,R$
that shares a non-basic variable with~$A$ is of the form
$q(\ldots,t_j,\ldots)$, with $t_j\preceq t_i$, and $A$ and $Q$ 
do not share any variable besides the ones in $\mathit{vars}(t_i) \cap \mathit{vars}(t_j)$.
Let $K_{1}\leftarrow c_{1},
B_{1},~\ldots,~K_{m}\leftarrow c_{m}, B_{m}$ be all clauses
of $\mathit{Cls}$ (with $\vars(\mathit{Cls})\cap\vars(C)=\emptyset$) 
such that, for \( i\eq 1,\ldots ,m
\), (i)~there exists a most general unifier $\vartheta_i$ of $A$ and
$K_i$, and (ii)~the constraint $(c, c_{i})\vartheta_i$ is
satisfiable.
Then, by unfolding $A$ in $C$ we derive the clauses
$(H\leftarrow c, c_{1},
{L}, B_{1}, {R})\vartheta_1,
\ldots,
(H\leftarrow c, c_{m}, {L}, B_{m},
{R})\vartheta_m$.
By Property~P2 $A$ is a linear atom,
and by the hypothesis that every clause in $\mathit{Cls}$
has a  disjoint, quasi-descending slice decomposition,
we have that, for $i=1,\ldots,m,$ for every atom $E$ in ${L}, B_{i}, {R},$
$\mathit{height}(E\vartheta_i) \lesseq \max (\{\mathit{height}(E),
\mathit{height}(A),\mathit{height}(K_i)\})$.
Thus, if Property~P4 holds for $C$, then it also holds for  the clauses
$(H\leftarrow c, c_{1},
{L}, B_{1}, {R})\vartheta_1,
\ldots,$
\mbox{$(H\leftarrow c, c_{m}, {L}, B_{m},
{R})\vartheta_m$}.
Property~P4 also holds for each clause derived by {\it Replace}  steps, and hence
it also holds for $D$, whose body consists of atoms
taken from the body of clauses derived by  {\it Unfold} and {\it Replace} steps.

Property~P5 holds for each clause $D$ initially in {\it Defs}
because the body of clause $D$ is the
set of atoms occurring in the body of a goal $G$ in {\it Gs}
for which Hypothesis (i) holds.
Now we prove that the following two properties, which generalize
Property~P5, hold for each clause~$E$ derived  by
an {\it Unfold} step:
in each sharing block in the body of $E$,
(P5.1)~every non-basic variable has at most $M$ occurrences, and
(P5.2)~there exist $K\lesseq M$ predicate arguments 
such that every non-basic variable with more than one occurrence
in the body of $E$, also occurs in one of those $K$ arguments.

Suppose that P5.1 and P5.2 hold for a clause $C$ of the form
\mbox{$H \leftarrow c, {L}, A, {R}$}, and we unfold $A$ in $C$.
Let $K_{1}\leftarrow c_{1},
B_{1},~\ldots,~K_{m}\leftarrow c_{m}, B_{m}$ be all clauses
of $\mathit{Cls}$ (with $\vars(\mathit{Cls})\cap\vars(C)=\emptyset$) 
such that, for \( i\eq 1,\ldots ,m
\), (i)~there exists a most general unifier $\vartheta_i$ of $A$ and
$K_i$, and (ii)~the constraint $(c, c_{i})\vartheta_i$ is
satisfiable.
Then, by unfolding $A$ in $C$ we derive the clauses~$C_1:(H\leftarrow c, c_{1},
{L}, B_{1}, {R})\vartheta_1,\
\ldots,\
C_m:(H\leftarrow c, c_{m}, {L}, B_{m},
{R})\vartheta_m$.
By Properties~P2 and~P3, and by the existence of
a disjoint, quasi-descending slice decomposition, for $i=1,\ldots,m,$ 
the number of occurrences of any variable with more than one occurrence
in the body of $C_i$, is not larger
than the maximal number of occurrences of any variable in the body of $C$, and hence
Property~P5.1 holds for the body of $C_i$.

Moreover, suppose that in every sharing block in 
the body of $C$ there exist $K\lesseq M$ arguments 
$t_1,\ldots,t_K$ such that every variable variable with more than one occurrence,
also occurs in one of those $K$ arguments.
By Property~P3, we may assume that $t_1,\ldots,t_K$ are maximal with respect to the 
$\preceq$ relation and do not share any variable.
Looking at the {\it Unfold} procedure, the atom 
$A$ selected for unfolding must have one among $t_1,\ldots,t_K$ as an argument, say $t_1$.
By our hypotheses, if by unfolding $A$ the argument $t_1$ is replaced by more than one term,
these new terms must appear in different sharing blocks, and hence the number
of maximal arguments in each sharing block does not increase.
Thus, Property P5.2 holds for $C_1,\ldots,C_m$.
Similarly, we can prove that Properties P5.1 and P5.2 hold for each clause 
derived by {\it Replace} steps, 
and hence Property~5 holds for~$D$, whose body consists of a sharing block
of a clause derived by  {\it Unfold} and {\it Replace} steps.

Now, from Properties~P4 and~P5 it follows that there exists an integer $J$,
depending on~$H$ and $M$, such that in the body of $D$ there are at most $J$ distinct
variables. Thus, there exists $N$ such that the body of $D$ has at most $N$ atoms of height
not larger than $H$, and hence the thesis holds.  \hfill $\Box$

\end{document}